\documentclass[11pt, oneside]{article}   	
\pdfoutput=1
\usepackage{jheppub}

\usepackage{amsmath, amsthm, graphicx,amssymb,subfigure}
\usepackage[all]{xy}
\preprint{DIAS-STP-16-13}

\newcommand{\bxo}{\bar{X}^{\rho \dot{\rho}}}
\newcommand{\xo}{X_{\rho \dot{\rho}}}

\newcommand{\tr}{\operatorname{tr}}
\newcommand{\Tr}{\operatorname{Tr}}

\allowdisplaybreaks

\author[]{Yuhma Asano,}
\author[]{Veselin G. Filev,}
\author[]{Samuel Kov\'a\v{c}ik,}
\author[]{Denjoe O'Connor,}
\affiliation[]{School of Theoretical Physics,\\ 
       Dublin Institute for Advanced Studies, \\
       10 Burlington Road, 
       Dublin 4, Ireland.}
       
\emailAdd{yuhma@stp.dias.ie}
\emailAdd{vfilev@stp.dias.ie}
\emailAdd{skovacik@stp.dias.ie}
\emailAdd{denjoe@stp.dias.ie}

\abstract{We study the second derivative of the free energy with
  respect to the fundamental mass (the mass susceptibility) for the
  Berkooz-Douglas model as a function of temperature and at zero 
  mass. The model is believed to be holographically dual to a D0/D4
  intersection. We perform a lattice simulation of the system at finite
  temperature and find excellent agreement with predictions from the
  gravity dual. }

\title{A Computer Test of Holographic Flavour Dynamics II}

\begin{document}

\maketitle

\section{Introduction}
Gauge/gravity duality \cite{Maldacena:1997re, Itzhaki:1998dd}, the
idea that gravity can capture the dynamics of strongly coupled gauge
theories and vice versa continues to fascinate theoretical
physicists. Numerous applications ranging from condensed matter
physics to heavy ion collisions have been proposed. Most of them
exploit the weak/strong coupling duality of the correspondence.
However, it is this property that makes the correspondence difficult to
test, especially in a non-supersymmetric setting. In this paper we
continue recent efforts to test a particular regime of the
correspondence at finite temperature using mainly lattice
simulations \cite{Anagnostopoulos:2007fw, Catterall:2008yz,
  Hanada:2008ez, Catterall:2009xn, Hanada:2013rga, Kadoh:2015mka, Filev:2015hia}. We focus on the Berkooz-Douglas (BD) matrix model \cite{Berkooz:1996is}, a flavoured version of the BFSS matrix model \cite{Banks:1996vh}, holographically dual to the D0/D4 system \cite{Karch:2002sh,
  Mateos:2007vn}.

In ref.~\cite{Filev:2015cmz} the lattice formulation of the BD matrix
model was studied. The model was studied both holographically and with
computer simulations focusing on the fundamental condensate of the
theory as a main observable. In the large $N$ limit, as the mass
parameter is varied, gauge/gravity duality predicts the existence of a meson
melting phase transition, corresponding to a topology change
transition in the supergravity set-up. The studies of
ref.~\cite{Filev:2015cmz} show a remarkable agreement between theory
and simulations in the deconfined phase of the theory. It was
speculated that in this phase there is a cancellation mechanism for
the $\alpha'$ corrections to the condensate. The studies were
conducted at two different temperatures and for a variety of bare
masses.

In this paper we consider the opposite regime studying the
susceptibility of the condensate with respect to the bare mass, at
vanishing bare mass and for a range of different temperatures. The
advantage of this approach is that at high temperatures the BD model
can be studied perturbatively \cite{High-T-BD} and at low
temperatures we have a gauge/gravity prediction. This allows us at high 
temperatures to verify the validity of our lattice approach against the perturbative 
results of ref.~\cite{High-T-BD} while at low temperatures to compare with 
the predictions of gauge/gravity duality. Furthermore, we find that if we go 
to sufficiently high order in perturbation theory we can extrapolate the high temperature
expansion to intermediate temperatures. If the cancellation mechanism
for the $\alpha'$ corrections to the fundamental condensate does take
place we may expect to obtain agreement of these extrapolated
high temperature  results with the gauge/gravity curve. Remarkably, the low 
temperature curve obtained from the D0/D4 holographic set-up and the high 
temperature expansion curves are indeed very close in the intermediate 
temperature regime $T\sim \lambda^{1/3} $. 

The paper is organised as follows. In section \ref{Mass susceptibility}, we briefly review the
lattice formulation of the BD matrix model and its high temperature
expansion. In section \ref{Holographic description}, with details in appendix \ref{derivation dc/dm}, we present the derivation of the slope of the
condensate from supergravity. In section \ref{Lattice Results}, we present our results for
the slope of the fundamental condensate. We conclude with a discussion
in section~\ref{Conclusion}.

\section{Mass susceptibility of the condensate at high temperature}
\label{Mass susceptibility}
The BD model in euclidean 1+0 dimensions is given by the following action \cite{VanRaamsdonk:2001cg, Filev:2015cmz, High-T-BD}:
\begin{align}
 S_E=N\int_0^\beta d \tau  \, 
 &\Bigg[
 \Tr\left( \frac{1}{2}D_\tau X^a D_\tau X^a 
 + \frac{1}{2}D_\tau \bar{X}^{\rho \dot{\rho}} D_\tau X_{\rho \dot{\rho}} 
 + \frac{1}{2} \lambda^{\dagger\rho} D_\tau \lambda_\rho 
 + \frac{1}{2} \theta^{\dagger\dot{\rho}} D_\tau \theta_{\dot{\rho}} \right) 
 \nonumber\\
 &+\tr \left( D_\tau \bar{\Phi}^\rho D_\tau \Phi_\rho 
 + \chi^\dagger D_\tau \chi \right) 
 \nonumber\\
 &-\Tr \left( 
 \frac{1}{4} [X^a,X^b]^2
 + \frac{1}{2}[X^a, \bxo][X^a,\xo]
 \right)
 \nonumber\\
 &+\frac{1}{2}\Tr \sum_{A=1}^3\mathcal{D}^A\mathcal{D}^A
 +\tr \left( \bar{\Phi}^\rho (X^a-m^a)^2 \Phi _\rho \right) 
 \nonumber\\
 &-\Tr \left(
 -\frac{1}{2} \lambda^{\dagger\rho} \gamma^a [X^a , \lambda_\rho] 
 + \frac{1}{2}\theta^{\dagger\dot{\rho}} \gamma^a [X^a , \theta_{\dot{\rho}}] 
 - \sqrt{2}i \varepsilon^{\rho \sigma} \theta^{\dagger\dot{\rho}}
 [ X_{\sigma \dot{\rho}}, \lambda_\rho] 
 \right) \nonumber\\
 &-\tr \left( \chi^\dagger \gamma^a (X^a-m^a) \chi 
 + \sqrt{2}i \varepsilon^{\rho \sigma} \chi^\dagger \lambda_\rho \Phi_\sigma 
 + \sqrt{2}i \varepsilon_{\rho \sigma} 
 \bar{\Phi}^\rho \lambda^{\dagger\sigma} \chi \right) \,
 \Bigg] ,
 \label{action}
\end{align}
where 
\begin{align}
 &\mathcal{D}^A
 =\sigma^{A\; \sigma}_{\rho}\left(
 \frac{1}{2}[\bar{X}^{\rho \dot{\rho}}, X_{\sigma \dot{\rho}}]
 -\Phi_\sigma \bar\Phi^\rho
 \right)\, ,
\label{ADHM_D}
\end{align}
and the covariant derivative $D_\tau$ acts on the fields of the fundamental multiplet, $\Phi_\rho$ and $\chi$, as $D_\tau\;\cdot=(\partial_\tau -iA)\;\cdot$ .
The trace of the colour $SU(N)$ is written as $\Tr$ while that of the flavour $SU(N_f)$
is denoted by $\tr$.  The diagonal matrices, $m^a$,
correspond to the transverse positions of the D4--branes. 

Note that the overall factor of $N$ in equation (\ref{action}) implies that the dimensionless parameter $\beta$ is 
related to the temperature via: $\beta^{-1} =T/\lambda^{1/3}$, where $\lambda = N\,g^2$ is the 't Hooft coupling. 
In the same way the dimensionless parameter $m^a = m^a_q/\lambda^{1/3}$, where $m^a_q$ is the bare mass in physical units. 

The fundamental condensate is defined as the variation of the free energy density with respect to the bare mass parameter $m^a$: 
\begin{align}\label{cond}
 \langle{\cal O}_m^a\rangle
 &=\frac{\partial}{\partial m^a}\left( -\frac{1}{\beta}\log Z \right)=\left\langle\frac{\partial}{\partial m^a} \frac{ S_E}{\beta} \right\rangle
 =\left\langle\frac{N}{\beta}\int_0^\beta d\tau \tr \left\{
 2\bar\Phi^\rho (m^a-X^a)\Phi_\rho +\chi^\dagger \gamma^a \chi
 \right\}\right\rangle .
\end{align}
In this paper we focus on the mass susceptibility of the condensate at vanishing mass. The mass susceptibility of the condensate is, by definition,
\begin{equation}
\langle {\mathcal C}^m \rangle =\frac{\partial^2}{{\partial m^a}^2}\left( -\frac{1}{\beta}\log Z \right)=
\frac{1}{\beta}\left\langle\frac{\partial^2S_E}{{\partial m^a}^2} -\left(\frac{\partial S_E}{\partial m^a}\right)^2\right\rangle + \frac{1}{\beta}\left\langle\frac{\partial S_E}{\partial m^a}\right\rangle^2\ ,
\label{Cm_def}
\end{equation}
which can be written as
\begin{equation}
\langle {\mathcal C}^m \rangle = \left\langle {\partial_{m^a} {\cal O}^a_m}\right\rangle -\beta\left\langle ({\cal O}^a_m)^2 - \langle{\cal O}^a_m\rangle^2\right\rangle\ ,
\label{Cm_gen}
\end{equation}
using ${\cal O}_m^a = \partial S_E/\partial m^a$.
Now if we substitute $S_E$ with the action (\ref{action}), for the
operator ${\mathcal C}^m$ we obtain\footnote{Note that the expression for ${\mathcal C}^m$ in ref.~\cite{High-T-BD} differs by a factor of $N$. Note also that the factor of $1/5$ in the reference 
compensates the summation over $a=1,\,\dots\,,5$ while, in this paper, $a$ is not summed over in \eqref{Cm_def}, \eqref{Cm_gen} and \eqref{condensate_susceptibility}}
\begin{equation}
{\mathcal C}^m = \frac{2N}{\beta}\int_0^\beta d\tau \tr \bar\Phi^\rho \Phi_\rho -\frac{N^2}{\beta}\left(\int_0^\beta d\tau \tr \left\{ -2\bar\Phi^\rho X^a\Phi_\rho +\chi^\dagger \gamma^a \chi \right\}\right)^2\, .
\label{condensate_susceptibility}
\end{equation}

Recently, the high temperature expansion of the BD model was considered in
ref.~\cite{High-T-BD} using expansion in Matsubara modes and standard perturbation
theory. The following expansion of the mass susceptibility was obtained:
\begin{equation}\label{CmHighT}
\frac{\langle{\mathcal C}^m \rangle}{N}=  \beta^{-\frac{1}{2}}\Xi_6 +\beta (\Xi_7+\Xi_8) +O(\beta^{\frac{5}{2}}) \ , 
\end{equation}
where the constants $\Xi_6, \Xi_7$ and $\Xi_8$ can be measured by
simulating the pure matrix model obtained in the $T\to\infty$ limit and
have been tabulated in ref.~\cite{High-T-BD}. We can now use the
lattice formulation of the BD model proposed in
ref.~\cite{Filev:2015cmz} to compare to the high temperature expansion
formula (\ref{CmHighT}). Before we continue with the lattice studies
of the susceptibility let us focus on the holographic description of
the model at low temperature.

\section{Holographic description at low temperature}
\label{Holographic description}
At low temperature the BD model is proposed to be dual to the D0/D4 holographic set-up\footnote{The D0/D4 set-up belongs to a large class of Dp/Dp+4--brane intersections exhibiting universal properties such as the presence of a meson melting phase transition. For more details look at refs.~\cite{Babington:2003vm, First-order, Mateos:2007vn, Albash:, Hoyos:2006gb} as well as ref.~\cite{Erdmenger:2007cm} for an extensive review.}. The most understood case that we will focus on is the so called quenched approximation, when the flavour D4--branes are in the probe approximation \cite{Karch:2002sh}. In the near horizon limit the D0--brane supergravity background is given by
\begin{eqnarray}\label{metric}
ds^2&=&-H^{-\frac{1}{2}}\,f\,dt^2+H^{\frac{1}{2}}\left(\frac{du^2}{f}+u^2\,d\Omega_8^2\right)\ ,\nonumber \\
e^{\Phi}&=&H^{\frac{3}{4}}\ ,~~~~~~C_0 =H^{-1}\ ,
\end{eqnarray}
where $H=\left(L/u\right)^7$ and $f(u) = 1-(u_0/u)^{7}$. Here $u_0$ is the radius of the horizon related to the Hawking temperature via $T ={7}/({4\,\pi\,L})\left({u_0}/{L}\right)^{{5}/{2}}$ and the length scale $L$ is given by  $L^7=15/2\,(2\pi\alpha')^5\,\lambda$, with $\lambda$ the 't Hooft coupling.

To introduce matter in the fundamental representation we consider the addition of $N_f$ probe D4--branes. In the probe approximation $N_f \ll N$, their dynamics is governed by the Dirac-Born-Infeld action:
\begin{equation}\label{DBI}
S_{\rm DBI} = -\frac{N_f}{(2\pi)^4\,\alpha'^{5/2}\,g_s}\int\,d^4\xi\,e^{-\Phi}\,\sqrt{-{\rm det}||G_{\alpha,\beta}+(2\pi\alpha')F_{\alpha,\beta}||}\ ,
\end{equation}
where $G_{\alpha,\beta}$ is the induced metric and $F_{\alpha,\beta}$ is the $U(1)$ gauge field of the D4--brane, which we will set to zero. Parametrising the unit $S^8$ in the metric (\ref{metric}) as 
\begin{equation}
d\Omega_8^2 = d\theta^2+\cos^2\theta\, d\Omega_3^2 + \sin^2\theta\, d\Omega_4^2
\end{equation}
and taking a D4--brane embedding extended along $t,\,u,\,\Omega_3$ with a non-trivial profile $\theta(u)$, we obtain (after Wick rotation)
\begin{equation}\label{DBI-Wick}
S_{\rm DBI}^E =\frac{N_f\,\beta}{8\,\pi^2\,\alpha'^{5/2}\,g_s}\int\, du\,u^3\cos^3\theta(u)\,\sqrt{1+u^2\,f(u)\,\theta'(u)^2}\ .
\end{equation}
The embedding extremising the action (\ref{DBI-Wick}) can be obtain by solving numerically the corresponding non-linear equation of motion. The AdS/CFT dictionary then relates the behaviour of the solution at large radial distance $u$ to the bare mass and condensate of the theory via~\cite{Karch:2002sh,Mateos:2007vn}
\begin{equation}\label{Expansion-sin(theta)}
\sin\theta =\frac{\tilde m}{\tilde u}+\frac{\tilde c}{\tilde u^3}+\dots\ ,
\end{equation}
where $\tilde u = u/u_0$ and the parameters $\tilde m$ and $\tilde c$ are proportional to the bare mass and condensate of the theory. Therefore, the mass susceptibility of the condensate at zero bare mass $\langle{\mathcal C}^m\rangle$  is proportional to
\begin{equation}\label{Cm_prop}
\langle{\mathcal C}^m \rangle \propto -\left(\frac{d\tilde c}{d\tilde m}\right)\Big |_{\tilde m = 0} =\frac{7\pi}{2}\frac{\csc(\pi/7)\,\Gamma({3/7})\,\Gamma({5/7})}{\Gamma({1/7})^2\,\Gamma({2/7})\,\Gamma({4/7})}\ .
\end{equation}
The last expression was obtained by using that small $\tilde m$ implies small $\theta$, and hence the equation of motion for $\theta$ can be linearised and solved analytically. We refer the reader to appendix \ref{derivation dc/dm} for more details. Combining equation (\ref{Cm_prop}) with the exact expressions for the mass and condensate in terms of $\tilde m$ and $\tilde c$ {\cite{Mateos:2007vn,Filev:2015cmz}:
\begin{eqnarray}
m &=& m_q/\lambda^{1/3} = \frac{u_0\,\tilde m}{2\pi\alpha'}=\left(\frac{120\,\pi^2}{49}\right)^{1/5}\left(\frac{T}{\lambda^{1/3}}\right)^{2/5}\,\tilde m\ , \nonumber \\
\langle {\cal O}_m\rangle &=&-\frac{N_f\,u_0^3}{2\,\pi\,g_s\,\alpha'^{3/2}} \,\tilde c=\left(\frac{2^4 \,15^3\,\pi^6}{7^6}\right)^{1/5}\,N_f\,N_c\,\left(\frac{T}{\lambda^{1/3}}\right)^{6/5}\,(-2\,\tilde c)\ ,
\label{dictionary}
\end{eqnarray}
 we obtain
\begin{equation}\label{Cm-hol}
\langle{\mathcal C}^m\rangle =14^{1/5}15^{2/5}\pi^{9/5}\frac{\csc(\pi/7)\,\Gamma(3/7)\,\Gamma(5/7)}{\Gamma(1/7)^2\,\Gamma(2/7)\,\Gamma(4/7)}N_f\,N_c\,\left(\frac{T}{\lambda^{1/3}}\right)^{4/5}\approx 1.136\,N_f\,N_c\,\left(\frac{T}{\lambda^{1/3}}\right)^{4/5}\ .
\end{equation}
Equation (\ref{Cm-hol}) is the holographic prediction for the mass susceptibility of the fundamental condensate, which  in the next section we are going to test on the lattice.

\section{Lattice Results}
\label{Lattice Results}
In this section we use the lattice formulation of the BD model
developed in ref.~\cite{Filev:2015cmz} to test both the high
temperature expansion curve (\ref{CmHighT}) and the holographic
prediction (\ref{Cm-hol}). Remarkably, the two curves are already very
close in the intermediate $T/\lambda ^{1/3} \sim 1$ temperature
regime (see figure \ref{fig:1}) suggesting that the $\alpha'$ corrections to the mass susceptibility are small.

As was shown in section \ref{Mass susceptibility}, if we start with the action $S_E$ in equation (\ref{action}) we arrive at equations (\ref{Cm_gen}) and (\ref{condensate_susceptibility}). However on the
lattice we have to substitute the corresponding lattice action
$S_{bos}+S_{ps.f}$ \cite{Filev:2015cmz} for $S_E$, where $S_{bos}$ is
the discretised bosonic action and $S_{ps.f}$ is the pseudo-fermionic
action, in which the fermions are represented (modulo a neglected phase)
by pseudo-fermionic bosonic fields and a complicated fermionic matrix 
\cite{Filev:2015cmz}. As a result equation (\ref{condensate_susceptibility}) is no longer valid; however equation (\ref{Cm_gen}) remains valid, provided one substitutes the condensate operator ${\cal O}_m^a$ with the corresponding lattice operator, obtained by differentiating the lattice action with respect to the mass parameter $m^a$. In addition, the first term in equation (\ref{Cm_gen}) involves a further derivative with respect to $m^a$, which complicates the analysis due to the more complex mass dependence of the pseudo-fermionic
action\footnote{We refer the reader to ref.~\cite{Filev:2015cmz} for more details on the differentiation of the pseudo-fermionic action $S_{ps.f}$.}. Nevertheless, the fact that the second term in (\ref{Cm_gen}) is the variance of the condensate
operator remains true on the lattice. This means that we can use the simulation data for 
the calculation of the condensate to measure the variance term while the first term in (\ref{Cm_gen}) has to be
calculated directly as an expectation value.

The computation of the condensate susceptibility is a nice consistency check
of our codes, since it involves also the second momentum of the
simulation data.  We used this method for temperatures in the range
$1\leq T/\lambda^{1/3} \leq 5$. For temperatures $T <\lambda ^{1/3}$,
we found that the critical slowing down, related to the absence of a gap
in the supersymmetric system, impedes the estimation of the variance.
This is why for such low temperatures we used that the
condensate is, to a very good approximation, linear near $m=0$ and
since it vanishes at $m=0$ one can approximate
\begin{equation}
\langle {\mathcal C}^m \rangle \approx \langle{\cal O}^a_m\rangle /m^a\ ,
\end{equation}
for small $m^a$. This method also has the advantage that it imposes
the vanishing of the condensate at vanishing mass (which is true by
symmetry) and lowers the numerical error. We used this method to
estimate the slope at temperature $T = 0.8\,\lambda ^{1/3}$.

In our simulations we used lattice points out of $\Lambda = 16$, 24, 32, 48 for
temperatures $T/\lambda^{1/3} \lesssim 1$. 
At higher temperatures (in the range $1\lesssim T/\lambda^{1/3} \leq 4$) 
we lowered $\Lambda$ to preserve roughly the same lattice spacing.
For temperatures higher than $T = 4\,\lambda ^{1/3}$ 
we used $\Lambda = 4$ lattice points. 
The rank of the gauge group was fixed at $N=10$ and we used one family of
flavours $N_f=1$ to minimise the ratio $N_f/N$ and improve the probe
approximation. Note that although on the gravity side we used the
probe approximation, the lattice simulation was dynamical
\cite{Filev:2015cmz}. Finally, for our parameters the high
temperature curve (\ref{CmHighT}) is given by \cite{High-T-BD}
\begin{equation}
\langle{\mathcal C}^m \rangle=  14.08 \left(\frac{T}{\lambda^{1/3}}\right)^{1/2} - 3.02\left(\frac{T}{\lambda^{1/3}}\right)^{-1}+O({T}^{-\frac{5}{2}}) \ .
\label{CmHighT-part}
\end{equation}
In figure \ref{fig:1} we present our main result. The red dashed curve
is the holographic curve (\ref{Cm-hol}), and the black dashed curve is the
high temperature curve (\ref{CmHighT-part}). 
The blue bars represent the lattice simulations based on the lattice formulation developed in ref.~\cite{Filev:2015cmz} 
while the red bars correspond to independent lattice simulations based on a different lattice discretisation\footnote{
The latter lattice simulations use lattice derivatives in the second-order method \cite{AOCprep}. 
}.
The results of these simulations agree very well.
The red error bar at $T=\lambda^{1/3}$ has been obtained by extrapolating to $\Lambda = \infty$ using simulations with $\Lambda = 16$, 24, 32 and 48 (see figure \ref{fig:2}). 
The extrapolated result of $\left.\langle{\mathcal C}^m \rangle\right|_{T=\lambda^{1/3}}$ by a linear function is $11.26 \pm 0.29$ and the one by a quadratic function is $11.33 \pm 0.57$; they perfectly agree with the AdS/CFT prediction~(\ref{Cm-hol}) for $T=\lambda^{1/3}$ and $N=10$: $\left.\langle{\mathcal C}^m \rangle\right|_{T=\lambda^{1/3}}=11.36$.

Overall, one can observe excellent agreement of the
lattice simulation and the high temperature curve even for temperatures as low
as $T=\lambda^{1/3}$. One can also observe excellent agreement with
holographic predictions at temperatures $T\sim\lambda^{1/3}$.
Remarkably, even the high temperature curve is
very close to the holographic curve in this regime. As mentioned earlier this suggests that the $\alpha'$ corrections 
to the mass susceptibility are indeed very small.
\begin{figure}[t] 
   \centering
   \includegraphics[width=5.5in]{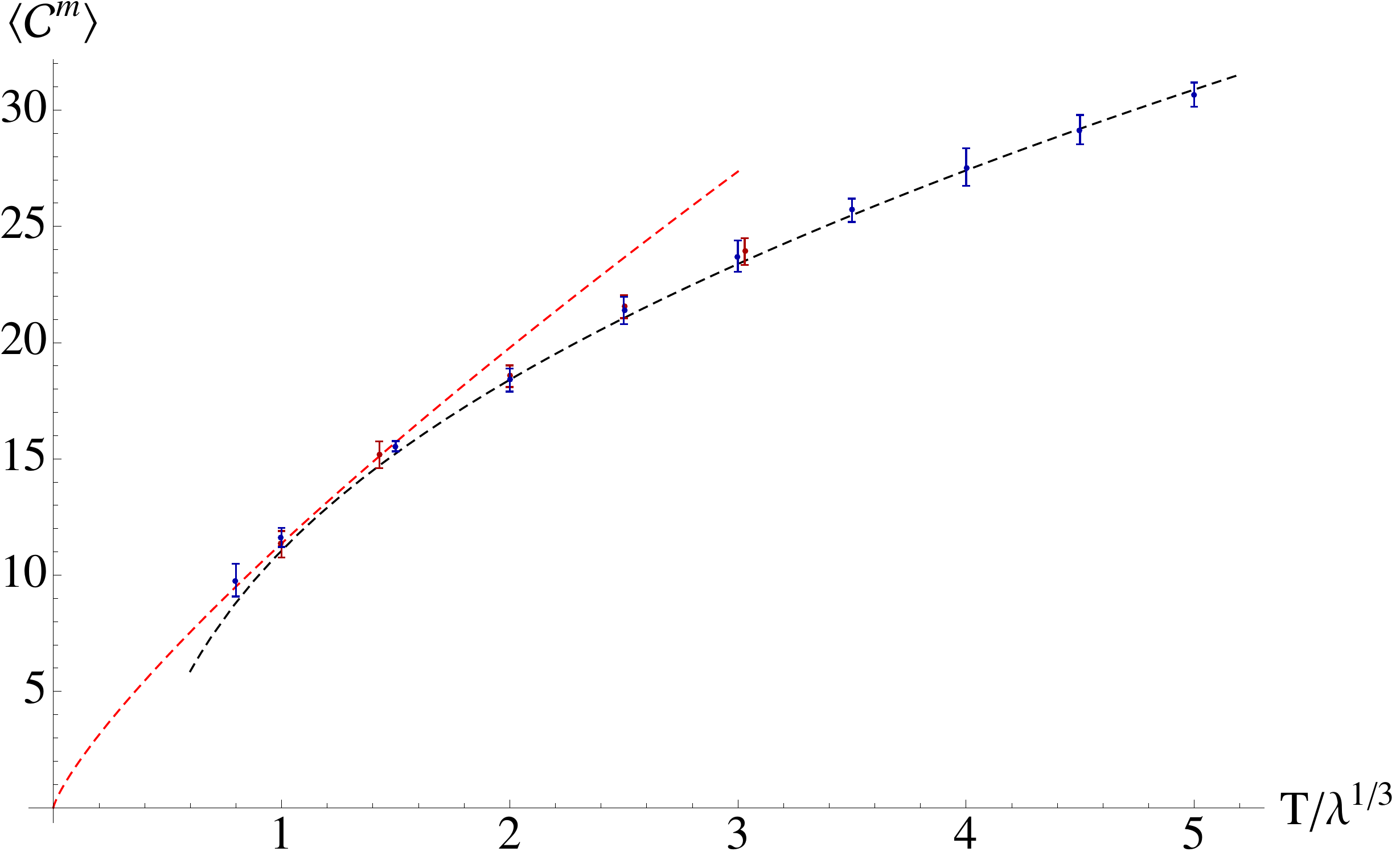} 
 \caption{\small The red curve represents the holographic prediction
   (\ref{Cm-hol}), while the black dashed curve corresponds to the
   high temperature expansion curve (\ref{CmHighT-part}). The blue bars
   represent the results of lattice simulations using the lattice discretisation in ref.~\cite{Filev:2015cmz}. The red bars correspond to independent lattice simulations based on a different lattice discretisation. }
   \label{fig:1}
\end{figure}
\begin{figure}[t] 
   \centering
   \includegraphics[width=5.5in]{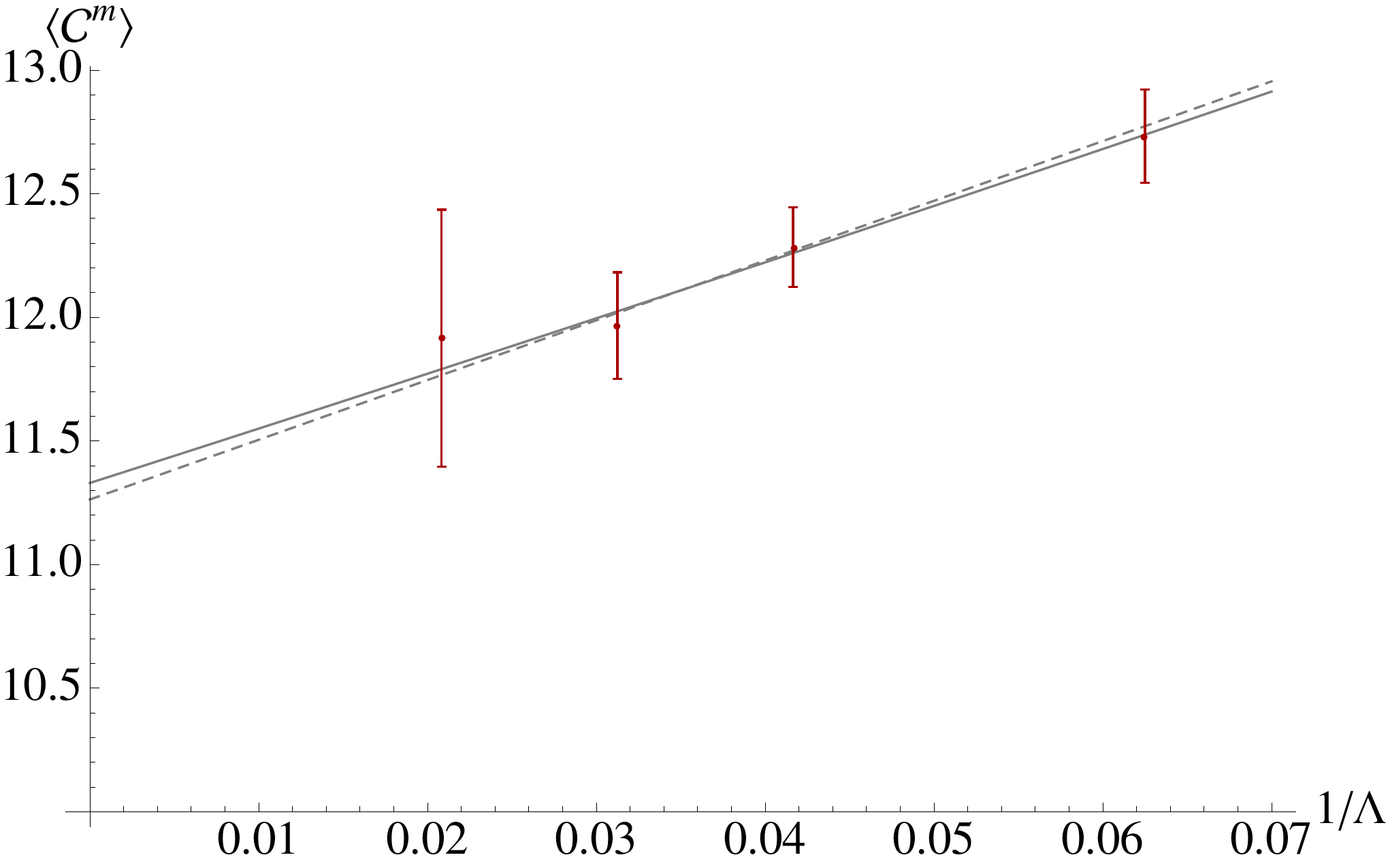} 
 \caption{\small The red bars correspond to measurements at $\Lambda = 16$, 24, 32 and 48 for $T=\lambda^{1/3}$ and $N=10$. The dashed gray line represents the linear extrapolation, and the solid gray curve corresponds to the quadratic extrapolation. 
 }
   \label{fig:2}
\end{figure}
\section{Conclusion}
\label{Conclusion}

In this paper we continue our investigation of the BD model and its
relation to the D0/D4--brane holographic set-up. The main
observable that we consider is the mass susceptibility of the fundamental
condensate at vanishing fundamental mass. We applied the recent
analysis of the high temperature regime of the BD matrix model to write
down a perturbative expression for the susceptibility at high
temperatures. We also review the holographic derivation of the
fundamental condensate and obtain an analytic result for the
susceptibility valid at low temperatures. Based on the observation of
ref.~\cite{Filev:2015cmz} that the $\alpha'$ corrections to the
condensate seem to be insignificant in the deconfined phase (which is
the relevant phase at vanishing bare mass) we expect to find good
agreement with the holographic curve not only at very low
temperatures, but also at intermediate temperatures $T\leq\lambda^{1/3}$.
Remarkably, the high temperature expansion curve also
remains valid down to intermediate temperatures $T\geq \lambda^{1/3}$ and is
in fact very close to the AdS/CFT curve.
Our lattice simulation is also in excellent agreement with both the high
temperature and low temperature predictions, verifying the
validity of the gauge/gravity correspondence.

Our results can be extended in several directions. The numerical
direction is to push the simulation to lower temperatures, higher rank
gauge groups (larger $N$) and larger $\Lambda$, the number of lattice points. 
The theoretical direction is to try to extend the validity
of the high temperature curve by considering higher order perturbation
theory such as in the studies of ref.~\cite{High-T-BD}.  Equally,
one can attempt to estimate the leading $\alpha'$ corrections to the
fundamental condensate. Such studies could potentially provide a more
rigorous test of the correspondence, which does not rely on lattice
simulations. It would also be satisfying to understand in more
details the suppression of the $\alpha'$ corrections in the black hole
(deconfined) phase of the D0/D4 system. Finally, one could attempt to
study corrections to the probe approximation by taking into account
the backreaction of the flavour D4--branes. We leave all of these
interesting directions for future work. 
\\ \\
{\bf Acknowledgments:} S.~K.~was supported by the Irish Research Council funding. The authors wish to acknowledge the Irish Centre for High-End Computing (ICHEC) for the provision of computational facilities and support (Project Name dsphy003c, dsphy004c and dsphy009c). The support from Action MP1405 QSPACE of the COST foundation is gratefully acknowledged. The work of V.~F. and D.~O. was supported in part by the Bulgarian NSF grant DN08/3.

\appendix

\section{Analytic expression for the condensate susceptibility}
\label{derivation dc/dm}
To obtain an expression for the slope of the condensate curve at vanishing bare mass, we will determine the mass dependence of the condensate at small masses. This corresponds to D4--brane embeddings entering the horizon at small angle $\theta_0$. Our strategy is to substitute $\theta(\tilde u) = \theta_0\,\eta(\tilde u)$ into the equation of motion for $\theta$, derived from the action (\ref{DBI-Wick}) and expand to leading order in $\theta_0$. This is equivalent to linearising the equation of motion for $\theta$. The resulting equation for $\eta$ is given by
\begin{equation}\label{eom-eta}
\eta''(u) +\frac{2+5\,u^7}{u^8-u}\eta'(u)+\frac{3\,u^5}{u^7-1}\eta(u) = 0\ ,
\end{equation}
where to simplify the notation we have replaced $\tilde u \to u$. The general solution of equation (\ref{eom-eta}) is given by
\begin{equation}
\eta(u) = C(1)\, {}_2F_1\left[{1}/{7}, \,{3}/{7}, \,{4}/{7}, \,u^7\right]+C(2)\, u^3{}_2F_1\left[{4}/{7}, \,{6}/{7}, \,{10}/{7}, \,u^7\right]\ .
\end{equation}
Imposing regularity at the horizon ($u=1$) fixes one of the integration constants and the solution regular at $u=1$ is given by
\begin{equation}\label{regular-eta}
\eta(u) = C(1)\,\left( {}_2F_1\left[{1}/{7}, \,{3}/{7}, \,{4}/{7}, \,u^7\right]- u^3\frac{\Gamma(4/7)^2\,\Gamma(6/7)}{\Gamma(1/7)\,\Gamma(3/7)\,\Gamma(10/7)}\,{}_2F_1\left[{4}/{7}, \,{6}/{7}, \,{10}/{7}, \,u^7\right]\right)\ .
\end{equation}
The remaining integration constant $C(1)$ can be fixed by imposing $\eta(1) =1$. However, we will not need its value to determine the dependence of the condensate $\tilde c$ on the bare mass $\tilde m$. Indeed, expanding equation (\ref{regular-eta}) at large $u$, we obtain
\begin{eqnarray}\label{eta-expand}
\eta(u)  = C(1)\left(\frac{\Gamma(2/7)\,\Gamma(4/7)}{\Gamma(3/7)^2}\,\frac{1}{u} -\frac{7\,\pi\csc(\pi/7)\,\Gamma(5/7)}{2\,\Gamma(1/7)^2\,\Gamma(3/7)}\frac{1}{u^3}+O\left(\frac{1}{u^5}\right)\right)\ .
\end{eqnarray}
Note that equation (\ref{Expansion-sin(theta)}) can be rewritten as
\begin{equation}\label{theta-reexpand}
\theta(u) = \frac{\tilde m}{u} +\frac{\tilde c}{u^3}+O(\tilde m^2)+O(1/u^5)\ ,
\end{equation}
where we have kept only the terms linear in $\tilde m$, since equation (\ref{eta-expand}) is valid only to a linear order in $\theta_0$. Therefore, to linear order in $\tilde m$ we have
\begin{eqnarray}
\tilde m =  C(1)\,\theta_0\,\frac{\Gamma(2/7)\,\Gamma(4/7)}{\Gamma(3/7)^2}; ~~~\tilde c = -C(1)\,\theta_0\,\frac{7\,\pi\csc(\pi/7)\,\Gamma(5/7)}{2\,\Gamma(1/7)^2\,\Gamma(3/7)} \, \ ,
\end{eqnarray}
and hence
\begin{equation}
-\left(\frac{d\tilde c}{d\tilde m}\right)\Big |_{\tilde m = 0} = -\lim_{m\to 0}\left(\frac{\tilde c}{\tilde m}\right) =\frac{7\pi}{2}\frac{\csc(\pi/7)\,\Gamma(\frac{3}{7})\,\Gamma(\frac{5}{7})}{\Gamma(\frac{1}{7})^2\,\Gamma(\frac{2}{7})\,\Gamma(\frac{4}{7})}\ ,
\end{equation}
which is the result used in equation (\ref{Cm_prop}).


\begin{thebibliography}{99}

\bibitem{Maldacena:1997re} 
  J.~M.~Maldacena,
  Int.\ J.\ Theor.\ Phys.\  {\bf 38}, 1113 (1999)
  [Adv.\ Theor.\ Math.\ Phys.\  {\bf 2}, 231 (1998)]
  doi:10.1023/A:1026654312961
  [hep-th/9711200].

\bibitem{Itzhaki:1998dd} 
  N.~Itzhaki, J.~M.~Maldacena, J.~Sonnenschein and S.~Yankielowicz,
  Phys.\ Rev.\ D {\bf 58}, 046004 (1998)
  doi:10.1103/PhysRevD.58.046004
  [hep-th/9802042].
  
\bibitem{Anagnostopoulos:2007fw} 
  K.~N.~Anagnostopoulos, M.~Hanada, J.~Nishimura and S.~Takeuchi,
  ``Monte Carlo studies of supersymmetric matrix quantum mechanics with sixteen supercharges at finite temperature,''
  Phys.\ Rev.\ Lett.\  {\bf 100}, 021601 (2008)
  [arXiv:0707.4454 [hep-th]].

\bibitem{Catterall:2008yz} 
  S.~Catterall and T.~Wiseman,
  ``Black hole thermodynamics from simulations of lattice Yang-Mills theory,''
  Phys.\ Rev.\ D {\bf 78}, 041502 (2008)
  [arXiv:0803.4273 [hep-th]].
  
\bibitem{Hanada:2008ez} 
  M.~Hanada, Y.~Hyakutake, J.~Nishimura and S.~Takeuchi,
  Phys.\ Rev.\ Lett.\  {\bf 102}, 191602 (2009)
  [arXiv:0811.3102 [hep-th]].
  
\bibitem{Kadoh:2015mka} 
  D.~Kadoh and S.~Kamata,
  ``Gauge/gravity duality and lattice simulations of one dimensional SYM with sixteen supercharges,''
  arXiv:1503.08499 [hep-lat].

\bibitem{Filev:2015hia} 
  V.~G.~Filev and D.~O'Connor,
  JHEP 1605 (2016) 167
  [arXiv:1506.01366 [hep-th]].

\bibitem{Catterall:2009xn}
  S.~Catterall and T.~Wiseman,
  JHEP {\bf 1004} (2010) 077
  [arXiv:0909.4947 [hep-th]].

\bibitem{Hanada:2013rga}
  M.~Hanada, Y.~Hyakutake, G.~Ishiki and J.~Nishimura,
  Science {\bf 344} (2014) 882
  [arXiv:1311.5607 [hep-th]].

\bibitem{Berkooz:1996is} 
  M.~Berkooz and M.~R.~Douglas,
  Phys.\ Lett.\ B {\bf 395}, 196 (1997)
  doi:10.1016/S0370-2693(97)00014-2
  [hep-th/9610236].
  
\bibitem{Banks:1996vh} 
  T.~Banks, W.~Fischler, S.~H.~Shenker and L.~Susskind,
  ``M theory as a matrix model: A Conjecture,''
  Phys.\ Rev.\ D {\bf 55}, 5112 (1997)
  [hep-th/9610043].
  
\bibitem{Karch:2002sh} 
  A.~Karch and E.~Katz,
  JHEP {\bf 0206}, 043 (2002)
  doi:10.1088/1126-6708/2002/06/043
  [hep-th/0205236].
  
\bibitem{Mateos:2007vn} 
D.~Mateos, R.~C.~Myers and R.~M.~Thomson,
  Phys.\ Rev.\ Lett.\  {\bf 97}, 091601 (2006)
  doi:10.1103/PhysRevLett.97.091601
  [hep-th/0605046].\\
  D.~Mateos, R.~C.~Myers and R.~M.~Thomson,
  JHEP {\bf 0705}, 067 (2007)
  doi:10.1088/1126-6708/2007/05/067
  [hep-th/0701132].
 
\bibitem{Filev:2015cmz} 
  V.~G.~Filev and D.~O'Connor,
  JHEP {\bf 1605}, 122 (2016)
  doi:10.1007/JHEP05(2016)122
  [arXiv:1512.02536 [hep-th]].
  
  \bibitem{High-T-BD}
  Y.~Asano, V.~G.~Filev, S.~Kov\'a\v{c}ik and D.~O'Connor,
  arXiv:1605.05597 [hep-th].
  
\bibitem{VanRaamsdonk:2001cg} 
  M.~Van Raamsdonk,
  JHEP {\bf 0202}, 001 (2002)
  doi:10.1088/1126-6708/2002/02/001
  [hep-th/0112081].
 
\bibitem{Babington:2003vm} 
  J.~Babington, J.~Erdmenger, N.~J.~Evans, Z.~Guralnik and I.~Kirsch,
  Phys.\ Rev.\ D {\bf 69}, 066007 (2004)
  doi:10.1103/PhysRevD.69.066007
  [hep-th/0306018].\\

\bibitem{Hoyos:2006gb} 
  C.~Hoyos-Badajoz, K.~Landsteiner and S.~Montero,
  JHEP {\bf 0704}, 031 (2007)
  doi:10.1088/1126-6708/2007/04/031
  [hep-th/0612169].
  
  \bibitem{First-order}
  I. Kirsch, ``Generalizations of the AdS/CFT correspondence,'' Fortsch. Phys. 52 (2004) 727 [arXiv:hep-th/0406274].
  
  
  \bibitem{Albash:}
  T.~Albash, V.~G.~Filev, C.~V.~Johnson and A.~Kundu,
  Phys.\ Rev.\ D {\bf 77}, 066004 (2008)
  doi:10.1103/PhysRevD.77.066004
  [hep-th/0605088].
  
\bibitem{Erdmenger:2007cm} 
  J.~Erdmenger, N.~Evans, I.~Kirsch and E.~Threlfall,
  Eur.\ Phys.\ J.\ A {\bf 35}, 81 (2008)
  doi:10.1140/epja/i2007-10540-1
  [arXiv:0711.4467 [hep-th]].

 \bibitem{AOCprep} 
  Y.~Asano and D.~O'Connor,
  ``Checking gauge/gravity duality with the BFSS model by different lattice discretisations,''
  in preparation.


\end{thebibliography}
\end{document}